\documentclass{webofc}
\usepackage[varg]{txfonts}   
\newcommand{\expe}[1]{{\sc{#1}}}

\def\psigal{\ifmmode{\psi}\else{$\psi$}\fi}
\def\avgpsigal{\ifmmode{\overline{\psi}}\else{$\overline{\psi}$}\fi}
\def\deltapsigal{\ifmmode{\Delta\psi}\else{$\Delta\psi$}\fi}
\def\errmax{3.96}\def\errmaxarcmin{237.7}
\def\errstd{1.24}\def\errstdarcmin{74.6}
\def\errgrd{0.27}\def\errgrdarcmin{15.9}
\def\errceb{0.22}\def\errcebarcmin{13.0}
\def\errctb{0.17}\def\errctbarcmin{10.2}
\def\errtbn{0.11}\def\errtbnarcmin{6.5}

\begin{document}
\title{Absolute calibration of the polarisation angle for future CMB B-mode experiments from current and future measurements of the Crab nebula}
%
%

\author{\firstname{J.} \lastname{Aumont}\inst{1}\fnsep\thanks{\email{jonathan.aumont@irap.omp.eu}} \and 
	\firstname{A.} \lastname{Ritacco}\inst{2} 
    \and
    \firstname{J. F.} \lastname{Mac\'{i}as-P\'erez}\inst{3}
    \and
    \firstname{N.} \lastname{Ponthieu}\inst{4}
    \and
    \firstname{A.} \lastname{Mangilli}\inst{1}       
}

\institute{IRAP, Universit\'e de Toulouse, CNRS, CNES, UPS, (Toulouse), France
\and
Institut de RadioAstronomie Millim\'etrique (IRAM), Granada, Spain
\and
Laboratoire de Physique Subatomique et de Cosmologie, Universit\'e Grenoble Alpes, CNRS, 53, av. des Martyrs, Grenoble, France
\and
Univ. Grenoble Alpes, CNRS, IPAG, 38000 Grenoble, France
          }

\abstract{%
A tremendous international effort is currently dedicated to observing the so-called primordial $B$ modes of the Cosmic Microwave Background (CMB) polarisation. If measured, this faint signal imprinted by the primordial gravitational wave background, would be an evidence of the inflation epoch and quantify its energy scale, providing a rigorous test of fundamental physics far beyond the reach of accelerators. At the unprecedented sensitivity level that the new generation of CMB experiments aims to reach, every uncontrolled instrumental systematic effect will potentially result in an analysis bias that is larger than the much sought-after CMB $B$-mode signal. The absolute calibration of the polarisation angle is particularly important in this sense, as any associated error will end up in a leakage from the much larger $E$ modes into $B$ modes. The Crab nebula (Tau A), with its bright microwave synchrotron emission, is one of the few objects in the sky that can be used as absolute polarisation calibrators. In this communication, we review the best current constraints on its polarisation angle from 23 to 353\,GHz, at typical angular scales for CMB observations, from WMAP, IRAM XPOL, Planck and NIKA data. We will show that these polarisation angle measurements are compatible with a constant angle and we will present a study of the uncertainty on this mean angle, making different considerations on how to combine the individual measurement errors. For each of the cases, the potential impact on the CMB $B$-mode spectrum will be explored.}
\maketitle
\section{Introduction}
\label{intro}

Cosmic Inflation, a paradigm for the first instants of the Universe that solves issues of the Big-Bang scenario and proposes a mechanism for the generation of the primordial perturbations, could have produced a background of gravitational waves. These gravitational waves would have imprinted a specific geometrical pattern in the polarisation of the Cosmic Microwave Background (CMB), the so-called $B$-mode primordial CMB signal, of an unknown amplitude scaled by a parameter $r$, the ratio between tensor and scalar primordial perturbations, related to the energy scale at which Inflation occurred (scalar perturbations give rise to $E$ modes only, and tensor perturbations to both $E$ and $B$ modes). A tremendous international effort is devoted to achieving this much sought-after measurement, which would test fundamental physics at energy scales beyond the reach of accelerators. The sensitivity of ground-based instruments is dramatically increasing as the number of detectors rises and satellite missions for the next decade are proposed. Goal sensitivities are announced to reach statistical detections of $r = 10^{-3}$. At these sensitivities, representing a two-orders-of-magnitude leap forward from present capabilities (the current upper limit is r<0.06 \citep{bicep}), the systematic effects will have to be understood and controlled to an unprecedented sensitivity. 
 
The absolute calibration of the polarisation angle is of a particular importance in this sense, as any associated error will end up in a leakage from the much larger $E$ modes into $B$ modes, leading to a potential analysis bias bigger than the coveted $B$-mode primordial signal. Indeed, a miscalibration of an angle $\Delta\psi$ is equivalent to a rotation of the polarisation plane by the same angle, mixing the true linear polarisation $Q$ and $U$ and consequently the polarised angular power spectra \citep[see e.g.][]{pagano}:\\[-26pt] 

\begin{eqnarray}\label{eq:mixing}
\tilde{C}_{\ell}^{EE} &=& C_{\ell}^{EE}\cos^2 (2 \Delta\psi) + C_{\ell}^{BB}\sin^2 (2 \Delta\psi) - C_{\ell}^{EB}\sin (4 \Delta\psi) ~\nonumber\\
\tilde{C}_{\ell}^{BB} &=& C_{\ell}^{BB}\cos^2 (2 \Delta\psi) + C_{\ell}^{EE}\sin^2 (2 \Delta\psi) + C_{\ell}^{EB}\sin (4 \Delta\psi) ~\nonumber \\
\tilde{C}_{\ell}^{TE} &=& C_{\ell}^{TE}\cos (2 \Delta\psi) - C_{\ell}^{TB}\sin (2 \Delta\psi)  ~\nonumber \\
\tilde{C}_{\ell}^{TB}&=& C_{\ell}^{TE}\sin (2 \Delta\psi) + C_{\ell}^{TB}\cos (2 \Delta\psi) ~\nonumber \\
\tilde{C}_{\ell}^{EB}&=& \frac{1}{2}\left(C_{\ell}^{EE}- C_{\ell}^{BB}\right)\sin (4 \Delta\psi) + C_{\ell}^{EB}\left(\cos^2 (2 \Delta\psi) - \sin^2 (2 \Delta\psi)\right),
\end{eqnarray}\\[-26pt]

\noindent where $\tilde{C}_\ell$ are the measured and $C_\ell$ are the true angular power spectra. In a given CMB observing frequency, this rotation is strictly identical to the effect of {\it cosmological birefringence}. Phenomena beyond the standard model of cosmology can cause cosmological birefringence: parity violations, either in the electromagnetic \citep[e.g. cosmological pseudo-scalar field, see e.g.][]{carroll} or in the gravitational sector \citep[e.g. chiral gravitational waves, see e.g.][]{lue}, or primordial magnetic fields \citep[see e.g.][]{planck_pmf}. Parity violating processes can also produce $C_\ell^{TB}$ and $C_\ell^{EB}$ signals, otherwise null in the parity conserving standard cosmological model, but no measurement evidence has been reported yet. Finally, CMB foregrounds, such as dust emission from our Galaxy can additionally produce non-zero $C_\ell^{TB}$ and $C_\ell^{EB}$ spectra \citep{pipxxx}.

In the following, we will discuss the different strategies available in order to perform the absolute calibration of the polarisation angle. We will focus on the use of the Crab nebula (Tau A) as an absolute calibrator, as it jointly allows to test the instrumental system during operations and to preserve the $C_\ell^{TB}$ and $C_\ell^{EB}$ signals to probe cosmic birefringence. Following \citep{crabcalib}, we will review the available mesurements of the Crab nebula polarisation angle, consider several assumptions that can be made in order to efficiently combine them as a calibration reference and finally discuss the potential impact on the recovered parameter $r$. 

\section{Absolute calibration of the polarisation angle for CMB experiments}

Several calibration strategies for the polarisation angle of a CMB experiment can be envisaged, each with advantages and disavantages:\\[-12pt] 

\noindent{\small1.}\ {\bf Ground calibration}: Refers to a mechanical calibration of the polarisation angle prior to the observations. Even if depending on the design of an experiment, the mechanical alignment of a system orientation is in principle very good.  Nevertheless, it probes the state of the system before operations and needs anyway to be assessed during operations due to potential thermal effects and environment changes \citep[see e.g.][for a review]{debernardis}.\\[-12pt]

\noindent{\small2.}\ {\bf External calibration source}: Several authors have proposed external artificial calibration sources that could be used for the absolute calibration of the polarisation angle by CMB experiments, for example on a stratospheric balloon \citep{polocalc} or on a satellite \citep{calsat}. It can in principle provide a very good accuracy and tests the system during operations but is very expensive and has never been done.\\[-12pt]

\noindent{\small3.}\ {\bf Self-calibration}: The {\it self-calibration} assumes that the CMB intrinsic $C_\ell^{TB}$ and $C_\ell^{EB}$ spectra are zero. It thus minimizes these quantities against the absolute calibration of the polarization angle. This method usually has a very good accuracy and probes the system during operations. The main caveat is that the experiment using this technique loses its ability to constain the potential physics of $C_\ell^{TB}$ and $C_\ell^{EB}$ spectra. Foreground contamination might also be an issue.\\[-12pt]

\noindent{\small4.}\ {\bf Sky calibration}: Calibration using a polarized sky reference source. This method has the advantage of probing the system during operations and of allowing to avoid priors on the $C_\ell^{TB}$ and $C_\ell^{EB}$ spectra. However, its accuracy can be limited by the knowledge on the reference source (accuracy of the ancillary data, frequency dependence, time variability, extended source, \dots).\\[-15pt]

In this document, we evaluate the case of the {\it sky calibration} strategy. The Crab nebula (Tau A) is the brightest polarized source in the microwave sky at a few arcminute angular scale and is usually considered as the best sky option for the absolute calibration of the polarisation angle for CMB experiments. It is a well known, plerion-type supernova remnant, observed from radio to X-rays, well characterized in the radio and millimetre regime by a single synchrotron power-law, both in intensity and polarisation \citep{ritacco}. It is highly polarized (more than 20\,\%, $\sim$10\,\% in a typical CMB experiment beam) and has a spatial extension of about 7$\times$5 arcmin. No evidence for dust polarization contamination inside CMB experiment beams in microwave frequencies has been identified for now. We now review the existing measurements of the Crab nebula polarisation angle in the microwave regime and exploit them to derive the achievable accuracy in calibrating CMB experiments. 

\section{Crab polarisation angle measurements}

A compendium of the Crab nebula polarisation angle measurement in Galactic coordinates $\psi$, from 23 to 353\,GHz is given in \citep{ritacco}. It additionally introduces the \expe{Nika} measurement at 150\,GHz and recomputes the \expe{Planck-HFI} angles (100, 143, 217 and 353\,GHz) in a improved analysis with respect to \citep{crablfi}. The Crab polarization angle $\psi$ values are reported in Table 1. of \citep{ritacco}, together with their associated statistical and systematic uncertainties. 

For \expe{Planck-HFI}, we consider several uncertainties associated to the measurement of the Crab polarisation angle. We refer to the pre-flight errors on the absolute calibration of the polarisation angle \citep{rosset} as the \emph{ground} calibration error. These absolute calibration errors were later refined at 100, 143 and 217\,GHz in \citep{pipxlvi} using $C_\ell^{TB}$ and $C_\ell^{EB}$ minimisation, for which no cosmological signal is expected in the abscence of parity violating processes. We refer to these errors as $TB$ and $EB$, respectively. No $TB$ and $EB$ error were assessed for the 353\,GHz channel, so that we will always assign this channel measurement with the \expe{Planck-HFI} ground uncertainty.

The reported Crab polarisation angle values are compatible with a constant angle from 23 to 353\,GHz of $\avgpsigal=-88.26^\circ\pm0.27^\circ$ (inverse-square-noise weighted average using the \emph{ground} error for \expe{Planck-HFI}).

\section{Combined uncertainty on the Crab polarisation angle and miscalibration bias}
\label{sec:combined}
\label{sect:measurements}

In order to use the Crab nebula submillimetre polarisation angle $\psi$ as an absolute angle calibrator for CMB measurements, we are interested in the constraints on its uncertainty $\Delta\psi$, assessed from the measurements presented in the previous Section. Given the relatively small number of measurements and the variety of instruments, observing conditions and data processing, there is no unique way to combine them all into a single result with a well defined uncertainty. We therefore propose and test several combinations of these measurements to assess the combined uncertainty $\Delta\psi$: \\[-12pt] 

\noindent {\small\textbullet}\ \texttt{max}: We do not assume that the Crab polarisation angle \psigal{} is constant from 23 to 353\,GHz and we take the combined error \deltapsigal{} as the maximum difference between the inverse-square-noise weighted mean \avgpsigal{} and an individual measurement. The combined error is in this \texttt{max} case $\deltapsigal=\errmax{}^\circ$ ($\errmaxarcmin{}$\,arcmin).\\[-12pt] 

\noindent {\small \textbullet}\ \texttt{stddev}: We do not assume that the Crab polarisation angle \psigal{} is constant from 23 to 353\,GHz. We assume that the error on its value is dominated by the inter-frequency variations and take the standard deviation among the individual measurements to be the combined error on the Crab polarisation angle. In this \texttt{stddev} case, the combined error is $\deltapsigal=\errstd{}^\circ$ ($\errstdarcmin{}$\,arcmin).\\[-12pt] 

\noindent {\small \textbullet}\ \texttt{cst-PlanckGround}: We assume that the Crab polarisation angle \psigal{} is constant between 23 and 353\,GHz. The combined error is thus taken as the error on the inverse-square-noise weighted mean, taking the \emph{ground} systematic error for \expe{Planck-HFI} (see Sect.~\ref{sect:measurements}). The combined error is in this case $\deltapsigal=\errgrd{}^\circ$ ($\errgrdarcmin{}$\,arcmin).\\[-12pt] 

\noindent {\small \textbullet}\ \texttt{cst-PlanckEB}: As for the \texttt{cst-PlanckGround} case, the Crab polarisation angle is assumed constant. The difference with the \texttt{cst-PlanckGround} case is that we use the $C_\ell^{EB}$ minimisation assessment of the systematic error for the 100, 143 and 217\, GHz \expe{Planck-HFI} channels. The resulting combined error in that case is $\deltapsigal=\errceb{}^\circ$ ($\errcebarcmin{}$\,arcmin).\\[-12pt] 

\noindent {\small \textbullet}\ \texttt{cst-PlanckTB}: Same as \texttt{cst-PlanckEB}, but with the $C_\ell^{TB}$ minimisation \citep{pipxlvi}. The resulting combined error is $\deltapsigal=\errctb{}^\circ$ ($\errctbarcmin{}$\,arcmin).\\[-12pt] 

\noindent {\small \textbullet}\ \texttt{cst-PlanckTB+future} Same as \texttt{cst-PlanckTB} but adding 2 future measurements points having each a total error $\Delta\psi_{\rm Gal}^{\rm future}=0.2^\circ$. The combined error, assuming a constant polarisation angle for the Crab is in this case $\deltapsigal=\errtbn{}^\circ$ ($\errtbnarcmin{}$\,arcmin).\\[-12pt]

\begin{figure*}
\centering
\includegraphics[height=0.30\textwidth]{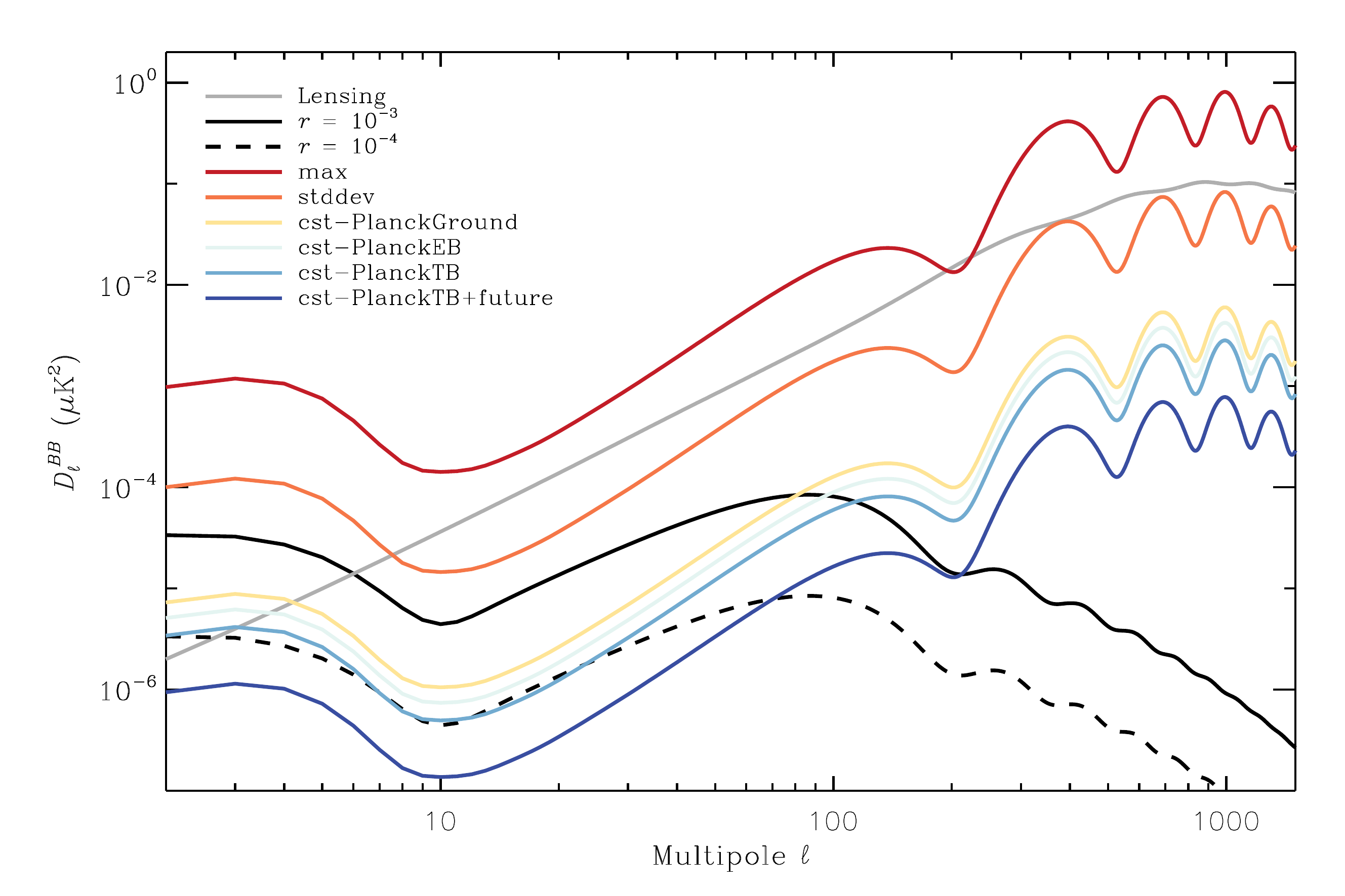}
\includegraphics[height=0.30\textwidth]{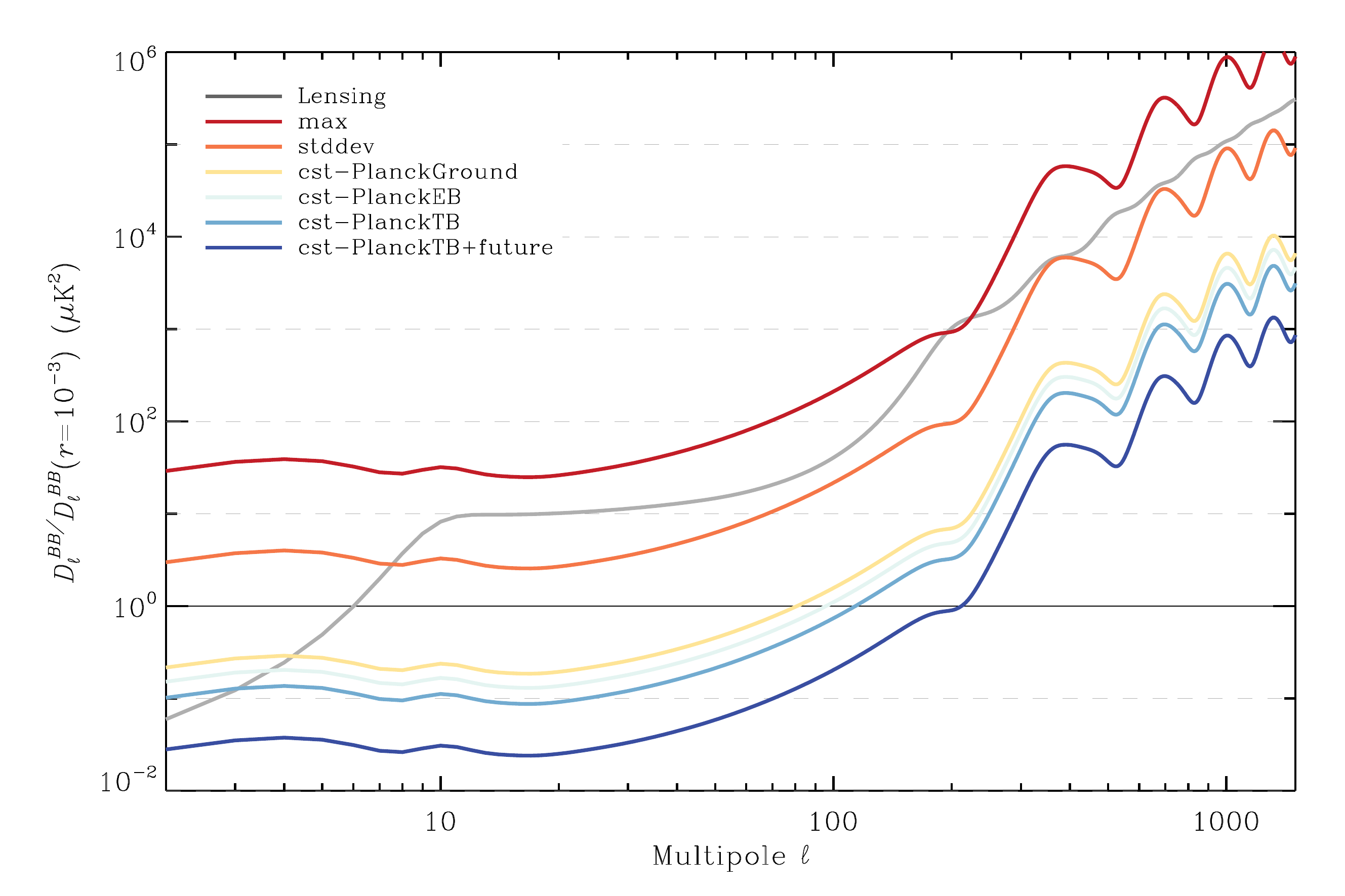}
\caption{\footnotesize {\bf Left panel}: $\Delta D_\ell^{BB}\equiv\ell(\ell+1)/(2\pi)\cdot\Delta C_\ell^{BB}$ power spectrum bias from $E$-$B$ mixing due to the mis-calibration of the absolute polarisation angle. This bias is plotted for the different absolute calibration errors \deltapsigal{} presented in Sect.~\ref{sect:measurements} (from red to blue, see legend). The $\Lambda$CDM best fit $D_\ell^{BB}$ primordial tensor model for $r=10^{-3}$ and $r=10^{-4}$ (solid and dashed black lines, respectively) and $D_\ell^{BB}$ lensing model (gray line) are also displayed. {\bf Right panel}: Same as left panel, but relative to the primordial tensor model for $r=10^{-3}$.}
\label{fig:bb_bias}
\end{figure*}

If one uses the Crab nebula as a calibrator, the uncertainty on its polarisation angle \deltapsigal{} sets a lower limit on the calibration error, and this has an impact on the magnitude of the corresponding $B$ modes bias, according to Eq.~\ref{eq:mixing}. Fig.~\ref{fig:bb_bias} shows the bias $\Delta C_\ell^{BB}\equiv\tilde{C}_\ell^{BB}-C_\ell^{BB}$ for the different combinations of experimental uncertainties presented above. We see that when we relax the assumption of a constant Crab polarisation angle from 23 to 353\,GHz (\texttt{max} and \texttt{stddev}), the spurious $B$-mode signal from $E$-$B$ mixing exceeds the primordial signal for $r=10^{-3}$ at \emph{all the angular scales}. If we assume the Crab polarisation angle to be constant (\texttt{cst-PlanckTB+future}, \texttt{cst-PlanckTB}, \texttt{cst-PlanckEB} and \texttt{cst-PlanckGround}), the biases range from $\sim3$ to $\sim30$\,\% of the primordial tensor signal for $\ell<10$, from $\sim20$ to more than $100$\,\% at $\ell\sim100$ and exceeds the signal in all cases for $\ell>250$.

\section{Likelihood analysis}
\label{sec:likelihood}

We quantify the effect of the absolute polarisation angle mis-calibration by looking at its impact on the recovery of the tensor to scalar ratio $r$ from CMB $B$-mode measurements. This is done in a likelihood analysis on the $r$ parameter, from simulated $\tilde{C}_\ell^{BB}$ measurements in the presence of a spurious signal $\Delta C_\ell^{BB}(\deltapsigal)$ coming from $E$-$B$ leakage due to the miscalibration of the polarization angle.

In each simulation, we consider a $\tilde{C}_\ell^{BB}$ measurement for $r=0$ and $\deltapsigal\neq0$, reading $\tilde{C}_\ell^{BB}=C_\ell^{BB,{\rm lens.}}+\Delta C_\ell^{BB}(\deltapsigal)$. The lensing only $C_\ell^{BB,{\rm lens.}}$ spectrum is computed from the \citep{planckcosmo} $\Lambda$CDM cosmology and the $\Delta C_\ell^{BB}(\deltapsigal)$ $E$-$B$ mixing component comes from Eq.~\ref{eq:mixing}. In each simulation, we draw randomly the \deltapsigal{} mis-calibration from a Gaussian distribution having a 1\,$\sigma$ dispersion corresponding to the error in each of the cases presented in Sect.~\ref{sec:combined}.

\begin{figure*}
\centering
\includegraphics[height=0.30\textwidth]{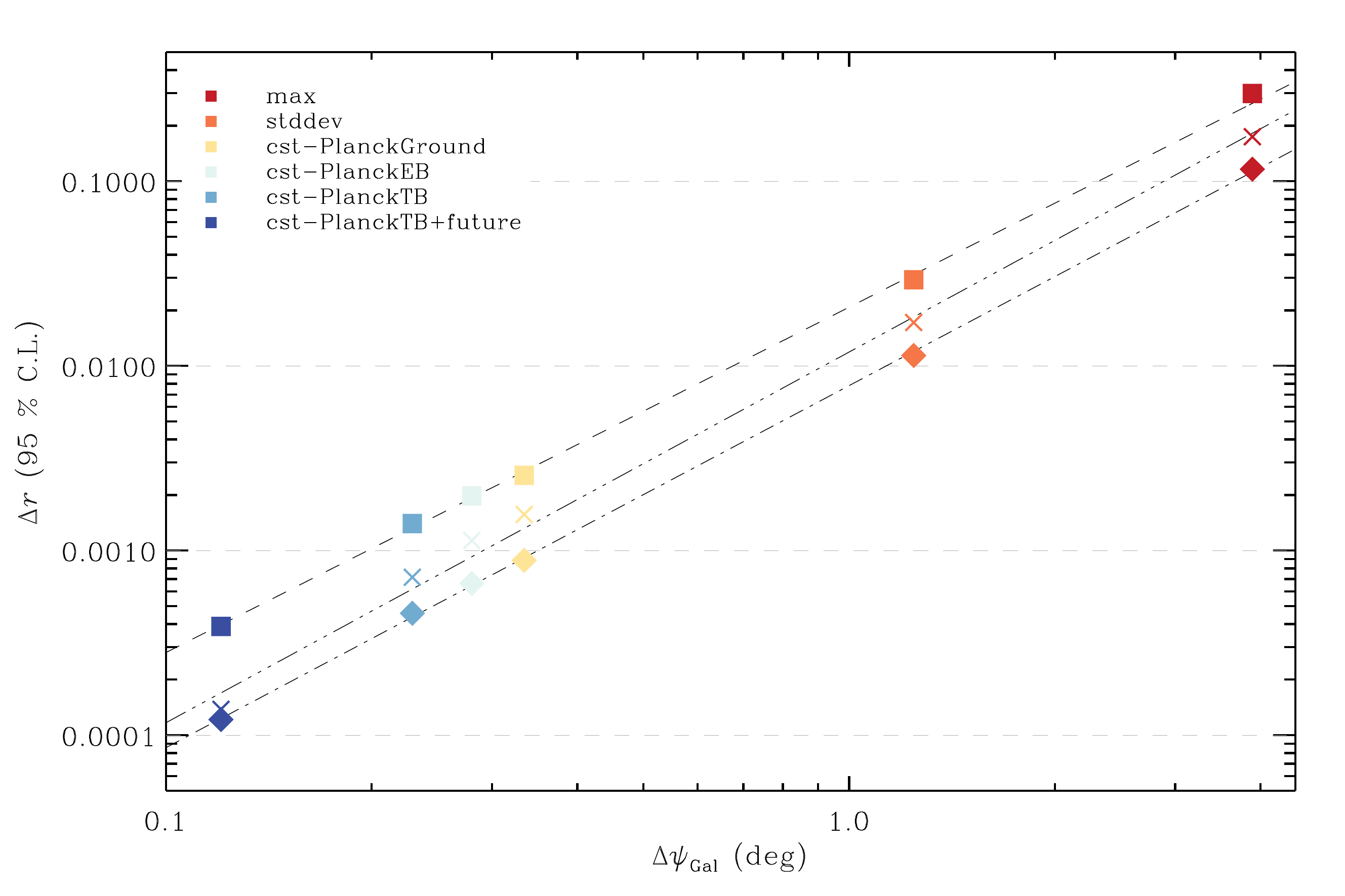}
\caption{\footnotesize Likelihood posterior on $r$ biases (with respect to an input signal of $r=0$) for the different cases of combined calibration errors (presented in Sect.~\ref{sect:measurements}) from 10\,000 Monte-Carlo simulations, as a function of the combined error on the angle $\Delta\psi_{\rm Gal}$ in degrees. They are computed independently for the recombination bump ($30<\ell<300$, squares), the reionisation bump ($2<\ell<30$, diamonds) and the combination of both ($2<\ell<300$, crosses).
\label{fig:posteriors}}
\end{figure*}

The likelihood function is computed on 10\,000 Monte-Carlo simulations. For each simulation, we build the posterior on $r$ and fit the bias $\Delta r$ with respect to $r=0$. We find the value $\Delta r(\rm 95\,\%\,C.L.)$ defined as the $r$ value for which 95\,\% of the simulations have a smaller $\Delta r$. This is done in 3 regimes of multipole range: for a typical ground-based experiment targetting the recombination bump ($\ell_{\rm min}=30$, $\ell_{\rm max}=300$), a satellite experiment with a large beam, having access to the reionisation bump only ($\ell_{\rm min}=2$, $\ell_{\rm max}=30$) and a satellite experiment having access to both the reionization and recombination bumps ($\ell_{\rm min}=2$, $\ell_{\rm max}=300$). 

The $\Delta r(\rm 95\,\%\,C.L.)$ values are presented in Fig.~\ref{fig:posteriors} for the recombination and reionisation bumps. We can see that the spurious $B$-mode polarisation coming from $E$-$B$ mixing is more penalising at high-$\ell$, resulting in higher $r$ biases for the recombination bump than for the reionisation bump {or both bumps together}. The two cases considered in Sect.~\ref{sec:combined} where we do not assume a spectrally constant polarisation for the Crab nebula (\texttt{max} and \texttt{stddev}) lead to biases on the $r$ posterior that are of the order of $r=10^{-2}$ or larger. In the cases where we assume that the Crab polarisation angle is constant (\texttt{cst-PlanckGround}, \texttt{cst-PlanckEB}, \texttt{cst-PlanckTB} and \texttt{cst-PlanckTB+future}), the biases on $r$ range from $r\sim10^{-4}$ to $r\sim3\times10^{-3}$. For the detection of $r=10^{-2}$, the best \emph{current} combined uncertainty on the Crab polarisation angle (\texttt{cst-PlanckTB} case) would lead to a potential 95\,\% C.L. bias of $\sim10$\,\% at the recombination bump and $\sim4$\,\% at the reionisation bump. With respect to $r=10^{-3}$, the current limits would lead to a 100\,\% bias at the recombination bump and 40\,\% at the lowest $\ell$ multipoles. Considering new measurements of the Crab polarisation angle, as in the \texttt{cst-PlanckTB+future} case, the bias could be shrunk down to negligible values for the measurement of $r=10^{-2}$ and down to $\sim10$ and $\sim30$\,\% of $r=10^{-3}$, for the reionisation and recombination bumps respectively.

\section{Conclusions}

The accurate absolute calibration of the polarisation angle of CMB experiments is a key element in the search for the CMB primodial $B$-modes and for CMB polarisation science beyond the standard model. In this communication, we present the current status of the calibration using the brightest polarized compact source in the microwave sky, the Crab nebula. We review the existing measurements that can be used to derive the polarisation angle of this source. We discuss the associated combined uncertainty and explore the potential bias on the measured $C_\ell^{BB}$ angular power spectrum. We first find that, in order to prevent biases larger than $r = 10^{-2}$, one has to assume the polarisation angle of the Crab nebula to be constant across microwave frequency. Under this assumption and with the current available measurements of the Crab polarisation angle, a CMB experiment targetting to measure $r=10^{-2}$ can use this source as an absolute calibrator. For the next generation of CMB experiment (e.g. \expe{LiteBIRD, CMB-S4, \dots}) more accurate measurements of the Crab nebula polarisation would be required in order to achieve the measurement of $r=10^{-3}$ using this calibration strategy.


%
%
%

\end{document}